# Real-space characterization of reactivity towards water at Bi$_2$Te$_3$(111) surface

Kai-Wen Zhang[1], Ding Ding[2], Chao-Long Yang[1], Yuan Gan[1], Shichao Li[1], Wen-Kai Huang[1], Ye-Heng Song[1], Zhen-Yu Jia[1], Xiang-Bing Li[1], Zihua Zhu[4], Jinsheng Wen[1,3], Mingshu Chen[2,*], Shao-Chun Li[1,3*]

1 National Laboratory of Solid State Microstructures, School of Physics, Nanjing University, Nanjing, Jiangsu 210093, China,

2 State Key Laboratory of Physical Chemistry of Solid Surfaces, Department of Chemistry, Xiamen University, Fujian 361005, China,

3 Collaborative Innovation Center of Advanced Microstructures, Nanjing University, Nanjing 210093, China,

4 Environmental Molecular Sciences Laboratory, Pacific Northwest National Laboratory, Richland, Washington 99352, United States

Email: scli@nju.edu.cn, chenms@xmu.edu.cn

Surface reactivity is important in modifying the physical and chemical properties of surface sensitive materials, such as the topological insulators (TIs). Even though many studies addressing the reactivity of TIs towards external gases have been reported, it is still under heavy debate whether and how the topological insulators react with H$_2$O. Here, we employ scanning tunneling microscopy (STM) to directly probe the surface reaction of Bi$_2$Te$_3$ towards H$_2$O. Surprisingly, it is found that only the top quintuple layer is reactive to H$_2$O, resulting in a hydrated Bi bilayer as well as some Bi islands, which passivate the surface and prevent from the subsequent reaction. A reaction mechanism is proposed with H$_2$Te and hydrated Bi as the products. Unexpectedly, our study indicates the reaction with water is intrinsic and not dependent on any surface defects. Since water inevitably exists, these findings provide key information when considering the reactions of Bi$_2$Te$_3$ with residual gases or atmosphere.

I. INTRODUCTION

Topological insulators (TIs) are characterized by a bulk band gap and robust metallic topological surface state (TSS) protected by time-reversal symmetry [1-3]. Owing to their topological properties, TIs have found great potential in both fundamental science and future applications [1-7]. Bismuth chalcogenide ($Bi_2X_3$), such as $Bi_2Se_3$ and $Bi_2Te_3$, is one type of the most extensively studied three dimensional TI materials, regarding to their TSS [8-11]. Meanwhile, Bismuth chalcogenide has also been widely explored as thermoelectric materials [12,13]. The surface stability upon exposing to air or residual gases is very crucial and has long been addressed, because the air or residual gases can significantly affect the surface states and thermoelectric performance [14-23]. A few groups reported the evolution of electronic properties of $Bi_2X_3$ upon exposure to atmosphere [18-21]. However, aging in ultrahigh vacuum was also observed but explained differently [24-26]. Upon exposure to CO, ARPES study showed the formation of quantum well states and Rashba splitting in $Bi_2X_3$ [27], but XPS experiments indicated there is no interaction [23]. Among the residual gases, $O_2$ and $H_2O$ are rather significant due to their inevitable existence in nature. Previous studies have shown that $Bi_2X_3$ was chemically inert to $O_2$ [20-23]. But, to date, the reactivity to water is still under debate. Even though quantum well states have also been observed upon exposure to water [16], XPS data show that $H_2O$ is not reactive to $Bi_2X_3$ [22]. The mechanism of how water reacts with $Bi_2X_3$ is still not clear.

In this study, we report the reaction of water with $Bi_2Te_3$(111) surface by using scanning tunneling microscopy (STM) and spectroscopy (STS). Both $Bi_2Te_3$ films and bulk single crystals exhibit the same phenomena. Even though XPS shows no prominent change of valence states, STM measurement indicates that surface morphology undergoes a drastic evolution upon exposure to water. Triangular shaped pits are formed in the top quintuple layer (QL) of the sample and the size of these pits grows with water dosage. Eventually the growing pits interconnect with each other until the full top QL is reacted. Finally, a surface terminated by hydrated Bi bilayers along with some Bi bilayer islands is formed and passivates the surface, preventing from further reaction. Spectroscopy measurements indicate the water acts as an effective n-type dopant. For comparison, exposure to oxygen gas has also been explored by STM, showing no prominent morphology changes except for some adsorbed features at surface defect locations. Since water and oxygen inevitably exists in air, the influence of water might be an important consideration in the reactions of $Bi_2Te_3$ with residual gases and air.

## II. EXPERIMENTAL DETAILS

The MBE growth and characterization of $Bi_2Te_3$ film were carried out in an ultrahigh vacuum (UHV) STM-MBE combined system (Unisoku) with a base pressure of $1 \times 10^{-10}$ Torr. The clean Si(111)-(7 × 7) surface was obtained by degassing at ~600 °C overnight following by cycles of flash annealing up to 1200 °C. High purity Bi (99.999%) and Te (99.999%) were evaporated from the standard Knudsen cells. RHEED was used to *in situ* monitor the film morphology. The Si substrate was kept at ~260 °C during $Bi_2Te_3$ growth. The ratio of Te/Bi flux was set to ~20:1, leading to a layer-by-layer growth mode. All the films used in this study have a thickness of more than 30 QLs. The bulk $Bi_2Te_3$ single crystals were grown using a self-flux method by melting the stoichiometric mixture of Bi and Te. The as-grown single crystals have nice cleavage surfaces. The single crystal $Bi_2Te_3$ was *in situ* cleaved in ultrahigh vacuum for STM measurement.

Characterizations of $Bi_2Te_3$ were carried out in UHV with a base pressure of $1 \times 10^{-10}$ Torr. STM images were acquired at both room temperature and ~100 K and STS spectrums were acquired at ~77 K. Constant current mode was adopted for STM scan. Tunneling spectrums were collected using a lock-in amplifier with a bias modulation of 10 mV at 1000 Hz. All STM images are processed by WSxM software [28]. Water source was purified by cycles of freeze-pump-thaw. The purity of Oxygen gas used in the experiment is better than 99.999%.

The XPS experiment was carried out in an UHV chamber equipped with an Omicron XPS (base pressure $5 \times 10^{-10}$ Torr). A load-lock chamber (base pressure $8 \times 10^{-9}$ Torr) was used for sample cleavage and $H_2O$ dosage. A monochromatized Al Kα X-ray source and a Sphera II Analyzer were used. XPS data were recorded in a perpendicular mode

## III. RESULTS AND DISCUSSION

$Bi_2Te_3$ takes the hexagonal layered structure stacked via interlayer van der Waals interaction. Each $Bi_2Te_3$ quintuple layer is constituted of five atomic layers of Te-Bi-Te-Bi-Te [8,9], see Fig. 1(a). Figure 1(b) shows the freshly cleaved $Bi_2Te_3$(111) surface terminated by the flat terraces with steps of ~1 nm high, consistent with the thickness of one quintuple layer. Figure 1(c) shows the atomic resolution image of freshly cleaved $Bi_2Te_3$(111) surface. When dosing water, e.g., to ~200 L, small pits and white protrusions are initially formed at the $Bi_2Te_3$ terrace, as seen in Fig. 2(a). The atomically resolved STM image, inset of Fig. 2(a), exhibits that the white protrusions reside right on top of the surface Te sites. Some of the white protrusions can hop from Te site to another at room temperature, see Fig. S1 in the Supplementary Material[29]. Further dosing of water induces more pits, and the pits can grow in lateral size. Figure 2(b) shows the surface after dosing water up to ~ 2200 L, where both the white protrusions and the pits are still observable. The pits are kept in triangular shape with the edges running along the high symmetrical directions of $Bi_2Te_3$(111). As further dosing water, such as to ~5500 L and ~7500 L, the growing triangular pits gradually interconnect with each other and form a fraction-like morphology, as shown in Figs. 2(c) and 2(d). We name the unreacted top terrace and the newly formed bottom terrace of the pits as I and II, respectively, as marked in Figs. 2(e) and 2(f). The apparent height measured from terrace I to II is ~0.5 nm and kept unchanged during the surface evolution. Besides the terraces I and II, small flat-top terraces are also formed next to the edge of some triangular pits, namely terrace III, as marked in Figs. 2(e) and 2(f). Terrace III is about 0.1 nm lower than the top-most terrace I, and ~0.4 nm higher than the terrace II. Upon dosing water, the percentage of terrace I decreases and those of terraces II and III are increased until the whole surface is terminated by terrace II with some islands of terrace III, as shown in Figs. 2(g) to 2(i).

Surprisingly, it is found that only the top QL is reactive to water. Once the full top QL is reacted, as shown in Fig. 2(e), the surface is no longer reactive to water, and no prominent evolution is observed for further water dosing, see also Fig. S2 in the Supplemental Material [29]. STM measurements on the $Bi_2Te_3$ thin films grown by MBE indicate the same behavior, as shown in Fig. S3 in the Supplemental Material [29].

Figure 3(a) shows the $dI/dV$ spectrum obtained on the unreacted terrace (terrace I) with various water dosage. For comparison, the $dI/dV$ curve obtained at the clean surface is also plotted. The position of spectrum minima (Min) and the conduction band minimum (CBM) can be easily distinguished, and assigned according to literature [30]. The d$I$/d$V$ in the gap region is linearly dependent on the bias which has been well ascribed to the contribution from TSS [30]. Both the Min and the CBM shift to a lower energy with increasing water dosage, i.e., the Fermi level of sample is shifted up towards the conduction band. This provides a strong evidence for an n-type doping effect. Figure 3(b) shows the quantitative evolution of the Min and the CBM as a function of water dosage, which is in agreement with previous report [16]. Such doping effect can be presumably induced by the adsorption of some resultant species on the terrace, e.g., the white protrusions observed on the terrace I. The linear component of the d$I$/d$V$ spectrum within the band gap region is kept almost unchanged all the spectrum measured at the water dosage of up to ~1760 L, suggesting that the TSS remains on the unreacted terrace area [16,22].

Figures 4(a) and 4(b) show the atomically resolved STM images of the unreacted region I and the newly formed region II respectively. Even though both of the regions exhibit the hexagonal symmetry and the lattice constants are nearly the same, the atomic-scale morphologies are rather different. The upper terrace shows slight electronic fluctuations in the STM contrast, while the lower terrace is populated with black point defects. Therefore, presumably we consider the newly formed terrace is not Te terminated. The atomically resolved terrace III is also hexagonal, see Figure 4(d). All the three terraces have nearly the same lattice constant and the same orientation. Figure S4 in the Supplemental Material shows the atomic registration of the three terraces, I, II, and III in the same region [29]. The electronic fluctuation observed at terrace I, despite the perfect atomic periodicity, might be induced by some adsorbed species at surface or subsurface.

Spectroscopy measurement gives more evidence that terraces I and II are different terminations. The STS taken at the terrace II is markedly different from the one taken at terrace I, as shown in Fig. 4(c). At first, the linear part of the spectrum at terrace II becomes less obvious which might indicate the dispersion change of TSS or weakened contribution from TSS [16,31,32]. Secondly, the valence band top and the conduction band bottom are more prominent and peak-like. According to the previous ARPES report, this might be due to the band quantization and the formation of quantum well states [16,27].

In contrast to the STM observation, XPS measurement showed no significant changes in the valence state of Bi and Te upon dosing water up to 120000 L, but the ratio of intensity for Bi to Te is increased upon dosing water, implying the top surface Te desorbs from the surface, see Fig. 5. In a previous study [22], single crystals were dosed with water and showed a significant intrinsic bulk doping. Therefore, it is more difficult to dope a doped crystals than an un-doped one, which might correspond to the reason why such an effect was not detected. Considering that the binding energy of 4f core level of Bi in Bi bilayer is rather similar to that in $Bi_2Te_3$ [33], it is not straightforward to conclude the reaction only via XPS measurements. It's worthwhile noting that the thickness of terrace III is ~0.4 nm, in line with the thickness of Bi bilayer, as reported previously [34,35]. Therefore a reasonable explanation is the formation of resultant Bi bilayer-related structure.

Based on STM / STS, and XPS experimental results, we adopt the similar reaction as proposed in the literature[16] that $H_2Te$ and hydrated Bi compounds are the products. At the initial stage, molecular $H_2O$ prefers to adsorb and diffuse across the $Bi_2Te_3$ terrace. Once gaseous $H_2Te$ is formed and desorbs from the surface at room temperature, a surface Te vacancy is formed, corresponding to the small pit observed by STM. On the other hand, the white protrusions observed by STM are considered as the $OH^-$ groups originating from $H_2O$ dissociation, which also correspond to the observed n-type doping effect to the $Bi_2Te_3$ surface.

The step edge height measurement in Fig. S5[29] indicates that the terrace II is composed of two atomic layers. Based on the chemical stoichiometry of $Bi_2Te_3$, this should be the hydrated Bi bilayer. The black holes in terrace II might correspond to the non-hydrated Bi sites or the Bi vacancies. In addition, formation of the Bi vacancies in such a hydrated bilayer provides the Bi atoms for the formation of terrace III. Another possibility might be that the $H_2O$ only catalyzes the Te to form $Te_2$. However, $Te_2$ dimer adsorbs at the $Bi_2Te_3$ surface at room temperature, and XPS measurements can also rule out this possibility by showing O peak when dosing $H_2O$ at room temperature [23].

Aiming to characterize the volatile product of $H_2Te$, low temperature STM measurement at ~100K has been performed during $H_2O$ adsorption. Figures 4(e) and 4(f) present the images for the same area before and after water dosing, respectively. After dosing water, we found the triangular pits start to form at the terrace and simultaneously some unknown structure is deposited at the step edge. These deposited structure at the step edge was not observed at room temperature. This observation can be explained by formation of the clustering of $H_2Te$ which would desorb from the surface at room temperature. Compared to the well-defined triangular shape formed at room temperature, the pits formed at low temperature are more irregular, as shown in Fig. 4(f). We believe at such low temperatures the pit doesn't gain enough thermal energy to ripen or coalesce.

Exposure to $O_2$ up to 10000 L has also been investigated, but no prominent surface morphology change was detected, except some oxygen related species adsorbed at the surface (Fig. S6)[29]. In contrast, the reactivity towards water is not dependent on the surface defects.

## IV. CONCLUSIONS

In summary, we have directly characterized the reaction of water at $Bi_2Te_3$(111) surface and found that only the top most QL of $Bi_2Te_3$(111) is reactive to water. The reacted surface is terminated mainly by hydrated Bi bilayer as well as some Bi islands. Such a hydrated Bi bilayer is inert to $H_2O$, and therefore preventing the second QL from further reaction. Such a reaction is not dependent on the population of surface defects. In contrast, $Bi_2Te_3$ surface is inert to $O_2$. This study gives at the first time the real-space reaction image, and is of importance in understanding the surface stability, in particular upon to residual gases, and helpful in TI based device applications.

## ACKNOWLEDGEMENTS

This work was supported by the State Key Program for Basic Research of China (Grants No.2014CB921103, No.2013CB922103), National Natural Science Foundation of China (Grants No.11374140, No.11374143), NCET-13-0282, and Open Research Fund Program of the State Key Laboratory of Low-Dimensional Quantum Physics.

Figure Captions:

**Figure 1.** (a) Crystal structure of $Bi_2Te_3$. (b) STM image of the cleaved surface of $Bi_2Te_3$ single crystal (Bias voltage $U$ = +1 V, Tunneling current $I_t$ = 100 pA; 100 nm × 100 nm). (c) the atomic resolution image of $Bi_2Te_3$ terrace ($U$ = +5 mV, $I_t$ = 250 pA; 5 nm × 5 nm).

**Figure 2.** (a-e) STM images obtaiend after dosing water of ~200 L, ~2200 L, ~5500 L, ~7500 L, ~12000 L, respectively ($U$ = +1 V, $I_t$ = 100 pA; 100 nm × 100 nm). (f) Line scan profile measured along the arrow in the inset which is extracted from (b) as the green rectangule. (g-i) STM apparent height distributions measured over (b)-(d).

**Figure 3.** (a) Differential conductance $dI/dV$ spectrum collected at the unreacted surface region of $Bi_2Te_3$ at various water dosage. The red and blue arrows indicate the positions of CBM and the minium, respectively. (b) Plots of CBM and the band minimum versus water dosage.

**Figure 4.** (a-b) Atomic resolution images of terraces I and II ($U$ = +5 mV, $I_t$ = 200 pA). The size is 6.0 nm × 4.5 nm (c) dI/dV spectrum obtained at terrace I and II. (d) Atomic resolution image of terrace III ($U$ = +5 mV, $I_t$ = 200 pA). The size is 2.8 nm × 2.8 nm. (e)-(f) STM images ($U$ = +1 V, $I_t$ = 100 pA; 215 nm × 100 nm) collected at the same area of $Bi_2Te_3$ surface before (a) and after (b) *in situ* water dosing.

**Figure 5.** XPS of $Te_{3d}$ (a) and $Bi_{4f}$ (b) of $Bi_2Te_3$ (111) and Bi/Te XPS ratio (c) with increasing $H_2O$ dosage.

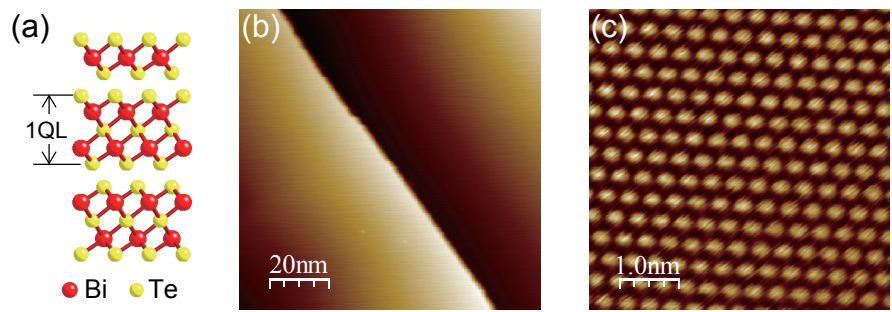

Figure 1

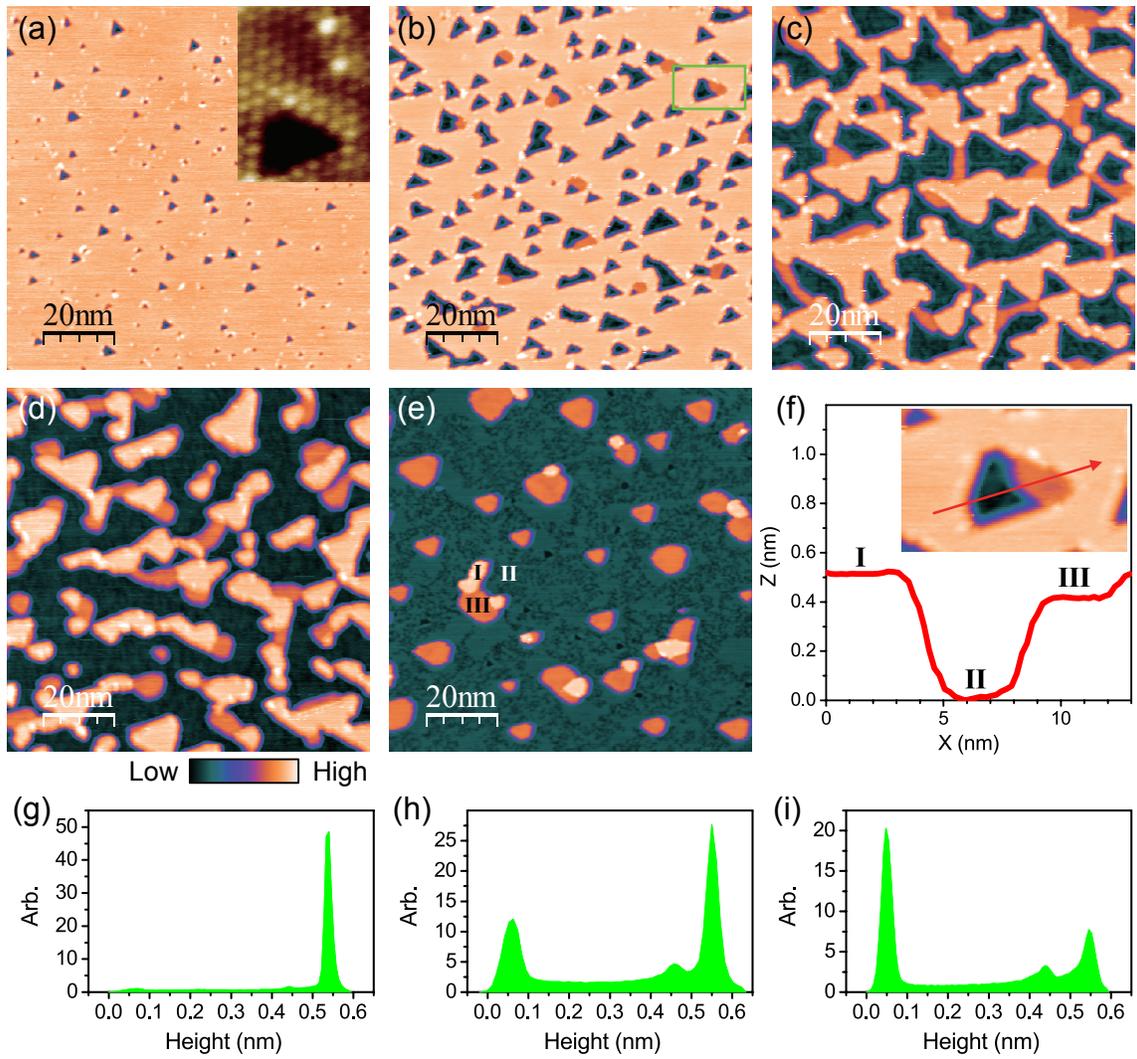

Figure 2

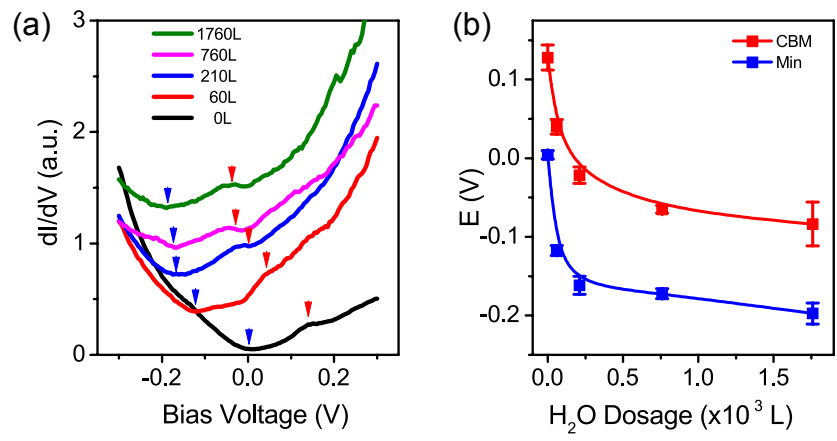

Figure 3

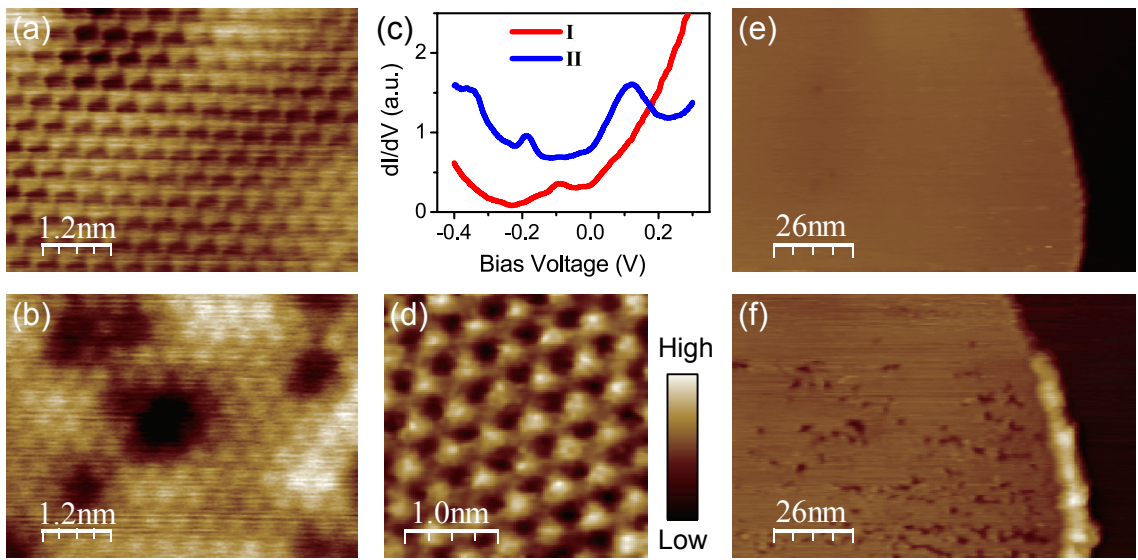

Figure 4

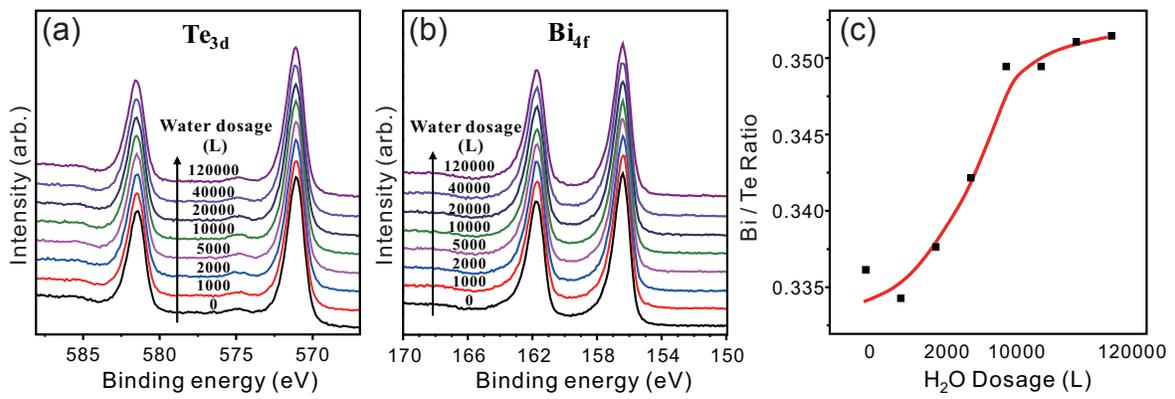

Figure 5